\documentclass[aps, prd, preprint, superscriptaddress, showpacs]{revtex4}
\usepackage{graphicx}
\usepackage{ulem}
\usepackage{epsfig}
\usepackage{color}
\usepackage{multirow}
\definecolor{My_red}        {cmyk}{0.00, 1.00, 1.00, 0.20}

\newcommand{\minitab}[2]{\begin{tabular}{@{}#1@{}}#2\end{tabular}}

\newcommand{\bmat}{\left(\begin{array}}
\newcommand{\emat}{\end{array}\right)}
\newcommand{\beq}{\begin{equation}}
\newcommand{\eeq}{\end{equation}}

\newcommand{\lsim}{\mathrel{\ltap}}

\usepackage[centertags]{amsmath}
\usepackage{amssymb}

\let\jnfont=\rm
\def\NPB#1,{{\jnfont Nucl.\ Phys.\ B }{\bf #1},}
\def\PLB#1,{{\jnfont Phys.\ Lett.\ B }{\bf #1},}
\def\EPJC#1,{{\jnfont Eur.\ Phys.\ Jour.\ C }{\bf #1},}
\def\PRD#1,{{\jnfont Phys.\ Rev.\ D }{\bf #1},}
\def\PRL#1,{{\jnfont Phys.\ Rev.\ Lett.\ }{\bf #1},}
\def\MPLA#1,{{\jnfont Mod.\ Phys.\ Lett.\ A }{\bf #1},}
\def\JPG#1,{{\jnfont J.\ Phys.\ G }{\bf #1},}
\def\CTP#1,{{\jnfont Commun.\ Theor.\ Phys.\ }{\bf #1},}
\def\JHEP#1,{{\jnfont JHEP \ }{\bf #1},}
\def\NPPS#1,{{\jnfont Nucl.\ Phys.\ Proc.\ Suppl.\ }{\bf #1},}
\def\CPC#1,{{\jnfont Computl.\ Phys.\ Commun.\ }{\bf #1},}
\def\CPL#1,{{\jnfont Chin.\ Phys.\ Lett. }{\bf #1},}
\def\APPB#1,{{\jnfont Acta\ Phys.\ Polon.\ B }{\bf #1},}

\def\lsim{\raise0.3ex\hbox{$<$\kern-0.75em\raise-1.1ex\hbox{$\sim$}}}
\def\gsim{\raise0.3ex\hbox{$>$\kern-0.75em\raise-1.1ex\hbox{$\sim$}}}

\begin{document}
\preprint{TTP13-004}

\title{Higgs decay to goldstini and its observability at the LHC}

\author{Tao Liu}
\affiliation{Institut f\"ur Theoretische Teilchenphysik, Karlsruhe
  Institute of Technology (KIT), D-76128 Karlsruhe, Germany}

\author{Lin Wang}
\affiliation{Institut f\"ur Theoretische Teilchenphysik, Karlsruhe
  Institute of Technology (KIT), D-76128 Karlsruhe, Germany}

\author{Jin Min Yang}
\affiliation{State Key Laboratory of Theoretical Physics,
      Institute of Theoretical Physics, Academia Sinica, Beijing 100190,
      China}

\begin{abstract}
If supersymmetry is broken independently in multiple sectors with
different scales, a number of goldstinos will be generated. One
linear combination of these goldstinos is massless and eaten by the
gravitino, while the orthogonal combinations acquire a tree level
mass and become the physical states named goldstini ($G'$). Compared
to the gravitino, such goldstini could couple more strongly to the
visible fields and lead to some exotic phenomenology. In this note
we first check the goldstini couplings in some GMSB models and find
that the goldstini-photon-neutralino interaction may be very small
while the goldstini-Z-neutralino and  goldstini-Higgs-neutralino
interactions may be sizable. This can induce a new decay mode for
the Higgs boson: $h\to G'+\chi \to Z+2G'$. Then in an effective
model with conservative fixed parameters we study the observability
of this decay channel at the LHC and find that it is not accessible
at the finished 8 TeV run (25 fb$^{-1}$) or 14 TeV run with 100
fb$^{-1}$, but might be observed at the high luminosity LHC
(14 TeV, 1000-3000 fb$^{-1}$) if the systematics of the backgrounds
can be well understood.
\end{abstract}

\maketitle

\section{Introduction}
Supersymmetry (SUSY) is a consistent extension of Poincar\'{e}
symmetry in quantum field theory and can solve the hierarchy problem
in the standard model due to the vanishing of quadratic divergence.
However, in a realistic particle theory, SUSY must be broken.
Usually it is assumed that SUSY breaking happens in a hidden sector
and then is transmitted to low energy fields via certain mechanisms.
This will generate a massless goldstino which acts as the longitude
component of gravitino in supergravity theory.
If SUSY breaking happens in only one sector, the interaction
of the gravitino is given by \cite{Fayet:1977vd,Dreiner:2008tw}
\begin{eqnarray}\label{one}
\mathcal{L}_{int}&=&\frac{1}{F}(\partial_\mu G^\alpha
J^\mu_\alpha+h.c.),\nonumber \\
J^\mu&=&\sigma^\nu\bar\sigma^\mu\psi D_\nu\phi^*-i\sigma^\mu\bar\psi
F_\phi+i\frac{1}{2\sqrt{2}}\sigma^\nu\bar\sigma^\rho\sigma^\mu\bar\lambda
F_{\nu\rho}+\frac{1}{\sqrt{2}}\sigma^\mu\bar\lambda D ,\label{eqn1}
\end{eqnarray}
where ($\phi$, $\psi$, $F_\phi$) are the boson, fermi and $\theta^2$
components of a chiral superfield; ($\lambda,A_\mu, D$) are
the fermi, gauge and $\theta^2\bar\theta^2$ components of a vector superfield;
$F_{\nu\rho}$ denotes the strength tensor of the gauge field $A_\mu$
and $D_\mu$ is the corresponding covariant derivative 
(throughout this paper we use the two component Weyl notation
for the fields).
The non-derivative couplings can be obtained after integrating by parts
and using the equation of motion. From Eq.(\ref{eqn1}) we see that
the gravitino does not play an important role in low energy phenomenology
unless the SUSY breaking scale $F$ is sufficiently low.

However, if SUSY is broken in multiple sectors, some exotic
phenomenology will be generated
\cite{Cheung:2010mc,Cheung:2010qf,Craig:2010yf,Cheng:2010mw,Mawatari:2011jy,Argurio:2011hs,Thaler:2011me,Cheung:2011jq,Argurio:2011gu}.
In such a scenario, each sector breaks SUSY independently at a scale
$F_i$ and gives a goldstino $\eta_i$. One linear combination of
$\eta_i$ is massless and eaten by the gravitino, while the
orthogonal combinations named goldstini acquire a tree level mass
$m_{G^\prime}=2m_{3/2}$ in SUGRA \cite{Cheung:2010mc} (possible loop
corrections to the goldstini mass have also been considered
\cite{Cheung:2010mc,Argurio:2011hs}). The true goldstino
(longitudinal component of gravitino) has an interaction in
Eq.(\ref{eqn1}) with $F=\sqrt{\sum_i F_i^2}$. Unconstrained by the
supercurrent, the interaction of goldstini can be quite different
and large enough to have intriguing phenomenology at colliders. For
example, in the framework of gauge mediated SUSY breaking (GMSB),
the goldstini can lead to some final states which are softer and
more structured at colliders \cite{Argurio:2011gu}. It can also
serve as a dark matter candidate \cite{Cheng:2010mw}. Additionally,
as noted in \cite{Thaler:2011me}, the goldstini may couple
'strongly' (much stronger than gravitino) with the lightest
observable-sector supersymmetric particle (LOSP), which then may
lead to some exotic decay channels for the Higgs boson, e.g., the
Higgs may decay to the LOSP plus a goldstini.

Given the importance and urgency of Higgs physics, any possible
exotic decays of the Higgs boson should be scrutinized. As is well
known, a Higgs-like particle around 125 GeV has been observed by
ATLAS and CMS collaborations \cite{125ATLAS,125CMS}, which motivated
numerous theoretical studies, especially the enhanced diphoton decay
has been intensively studied in various new physics models, such as
the low energy SUSY \cite{susy}, the little Higgs model \cite{lht},
the two-Higgs-doublet model \cite{2hdm}, the Higgs triplet model
\cite{triplet}, the models with extra dimensions \cite{extrad} and
other Higgs extensions \cite{hrrmodel}. In this note we focus on the
effects of goldstini and examine the exotic Higgs decay mode $h\to
\chi+G'$ followed by $ \chi \to Z+G'$ under the condition
$m_Z<m_\chi<m_h$ (here $\chi$ is the ordinary lightest neutralino
which is assumed to be the LOSP, $G^\prime$ denotes the goldstini).

We will make
a brief review on goldstini and explain why we focus on this channel
in Section II. Then we take a model-independent way to study the
Higgs decay $h\to \chi+G'$ followed by $ \chi \to Z+G'$
at the LHC in Section III. Finally,
we give our conclusion in Section IV.

\section{Theoretical motivations}
\subsection{A brief review on goldstini}
 For simplicity we assume that there are only two
sequestered sectors which break SUSY spontaneously. Following the
arguments in \cite{Thaler:2011me}, each of them can be parameterized
in a non-linear way:
\begin{eqnarray}
X_i=\frac{\eta_i^2}{2F_i}+\sqrt{2}\theta\eta_i+\theta^2F_i,
\end{eqnarray}
where $\eta$ is the so-called goldstino. Due to the
non-renormalization theorem of superpotential, visible sector only
obtains SUSY breaking information through non-trivial K$\ddot{a}$hler
potential $K$ and gauge kinetic functions $f$. After integrating out
hidden sector fields, $X_i$ couple to single species visible fields
$\Phi$ as
\begin{eqnarray}
K&=&\Phi^+\Phi\sum_i\frac{m_{\phi,i}^2}{F_i^{2}}X_i^+X_i,\\
f_{ab}&=&\frac{1}{g_a^2}\delta_{ab}\left(1+\sum_i\frac{2m_{a,i}}{F_i}X_i\right),
\end{eqnarray}
where $g_a$ denote the gauge coupling constants, $m_{\phi,a}$ are
respectively the soft masses for the scalar of the chiral superfield
and gauginos.
The trilinear and bilinear soft terms between multiple fields could also
be constructed, but since they are subleading and model-dependent in the
contribution to interactions between goldstino and visible sector,
we do not consider them here. Substituting the expression of $X_i$
into the above formula gives the Lagrangian up to order $1/F_i$:
\begin{eqnarray}
\mathcal
{L}=&-&\sum_im_{\phi,i}^2\phi^*\phi
+\sum_i\frac{m_{\phi,i}^2}{F_i}\eta_i\psi\phi^* \nonumber\\
~~~~&-&\frac{1}{2}\sum_i
m_{a,i}\lambda^a\lambda^a
-\sum_i\frac{im_{a,i}}{\sqrt{2}F_i}\eta_i\sigma^{\mu\nu}\lambda^aF^a_{\mu\nu}
+\sum_i\frac{m_{a,i}}{\sqrt{2}F_i}\eta_i\lambda^aD^a\label{gold}.
\end{eqnarray}
The mass eigenstates can be obtained by a rotation of $\eta_i$:
\begin{align}
G=\cos\theta\eta_1+\sin\theta\eta_2, \quad
G^\prime=-\sin\theta\eta_1+\cos\theta\eta_2,
\end{align}
where $\theta$ is defined by $\tan\theta=F_2/F_1$. It is easy to see
that the Weyl spinor $G$ is related to SUSY breaking scale
$F=\sqrt{F_1^2+F_2^2}$ while $G^\prime$ vanishes in the non-linear
form. Then the interaction Lagrangian becomes
\begin{eqnarray}
\mathcal{L}_G&=&\frac{m_\phi^2}{F}G\psi\phi^*
-\frac{im_a}{\sqrt{2}F}G\sigma^{\mu\nu}\lambda^aF^a_{\mu\nu}+
\frac{m_a}{F}G\lambda^aD^a  \label{LG},\\
\mathcal{L}_{G^\prime}&=&\frac{\tilde{m}_\phi^2}{F}G^\prime\psi\phi^*
-\frac{i\tilde{m}_a}{\sqrt{2}F}G^\prime\sigma^{\mu\nu}\lambda^aF^a_{\mu\nu}+
\frac{\tilde{m}_a}{F}G^\prime\lambda^aD^a \label{Leta},
\end{eqnarray}
with the parameter $m$ and $\tilde{m}$ defined as
\begin{align}
m_{\phi/a}=m_{\phi/a,1}+m_{\phi/a,2}, \quad
\tilde{m}_{\phi/a}=-m_{\phi/a,1}\tan\theta+m_{\phi/a,2}\cot\theta.
\end{align}
$\mathcal{L}_G$ is just the non-derivative version of
Eq.(\ref{eqn1}) and the couplings are proportional to the soft
masses as expected. For goldstini, their interactions with visible
fields could be enhanced as long as $\tilde{m}$ is larger than $m$.
Note that if $D^a$ get a vacuum expectation value, there will be
a mixing between $\lambda^a$ and $\eta_i$, which will give rise to
some special interactions. In the scenario with approximate
vanishing $\tilde{m}_a$ which we will discuss later,
such a mixing can be safely neglected in the interaction of $G^\prime$.

The mass of goldstini is given at tree level by
$m_{G^\prime}=2m_{3/2}$ due to the intrinsic property of SUGRA.
Additionally, there could be corrections to such a tree-level mass,
which, however, are model-dependent
\cite{Cheung:2010mc,Argurio:2011hs}. In our analysis we assume that
both $m_{G}$ and $m_{G^\prime}$ are much smaller than the Higgs
mass.

\subsection{Couplings between goldstini and Higgs boson}
Without the extra goldstini, the interaction of goldstino ($G$) with
the lightest neutralino ($\chi$) is given by
\begin{align}
\mathcal{L}=\frac{y_1}{\sqrt{2}F}\chi
Gh+\frac{y_2}{2\sqrt{2}F}\chi\sigma^{\mu\nu}
GF_{\mu\nu}^{\gamma}+\frac{y_3}{2\sqrt{2}F}\chi\sigma^{\mu\nu}
GZ_{\mu\nu}+\frac{y_4}{\sqrt{2}F}\bar\chi\bar\sigma^{\mu}GZ_\mu+h.c.
\end{align}
where $F^\gamma_{\mu\nu}$ is the photon field strength, $Z_\mu$ and
$Z_{\mu\nu}$ are respectively the $Z$-boson field and its field strength.
In the minimal supersymmetric standard model (MSSM), the above
parameters $y_{1,2,3,4}$ are given by
\begin{eqnarray}
y_1&=&-N_{11}^{-1}m_1m_Z\sin\theta_W\sin(\alpha+\beta)
+N_{21}^{-1}m_2m_Z\cos\theta_W\sin(\alpha+\beta)\nonumber\\
&&+N_{31}^{-1}(B_\mu\cos\alpha-m_{H_d}^2\sin\alpha)
+N_{41}^{-1}(-B_\mu\sin\alpha+m_{H_u}^2\cos\alpha),\\
y_2&=&-2im_\chi(N_{11}^*\cos\theta_W+N_{12}^*\sin\theta_W),\\
y_3&=&-2im_\chi(-N_{11}^*\sin\theta_W+N_{12}^*\cos\theta_W)
-2im_Z(N_{13}^*\cos\beta-N_{14}^*\sin\beta),\label{ZG1} \\
y_4&=&im_Z^2(-N_{11}^*\sin\theta_W+N_{12}^*\cos\theta_W)
+im_Zm_\chi(N_{13}^*\cos\beta-N_{14}^*\sin\beta) ,\label{ZG2}
\end{eqnarray}
where $N_{1i}$ denote the mixing between the lightest neutralino
and the gauginos or higgsinos.
We see that unless the SUSY breaking scale $F$ is small
enough, such interactions of goldstino
could hardly affect the 125 GeV Higgs.

Naively speaking, the interactions of the goldstini can be obtained
from the above expression with the goldstino $G$ replaced by the
goldstini $G'$ and each soft mass replaced by its corresponding
$\tilde{m}$. Then it is clear that for $\tilde{m}\gg m$ the
interactions of the goldstini can be significantly stronger than the
goldstino. However, there are some subtle differences which deserve
attention.

Next, we scrutinize the concrete low energy interactions of the
goldstini. The first one is its interaction with Higgs and
neutralino: $\frac{\tilde{y}_1}{\sqrt{2}F}\chi G^\prime h$. From
Eq.(\ref{Leta}) we can obtain the coefficient of this interaction:
\begin{align}
\tilde{y}_1=-N_{11}^{-1}\tilde{m}_1m_Z\sin\theta_W\sin(\alpha+\beta)
+N_{21}^{-1}\tilde{m}_2m_Z\cos\theta_W\sin(\alpha+\beta)\nonumber\\
+N_{31}^{-1}(\tilde{B}_\mu\cos\alpha-\tilde{m}_{H_d}^2\sin\alpha)
+N_{41}^{-1}(-\tilde{B}_\mu\sin\alpha+\tilde{m}_{H_u}^2\cos\alpha).
\end{align}
Similarly, the interaction with photon and neutralino takes a
form of $\frac{\tilde{y}_2}{2\sqrt{2}F}\chi\sigma^{\mu\nu}G^\prime
F_{\mu\nu}^\gamma$ with
\begin{align}\label{new}
\tilde{y}_2= -2im_\chi\left[\frac{\tilde{m}_{1}}{m_{1}}
N_{11}^*\cos\theta_W+\frac{\tilde{m}_2}{m_2}N_{12}^*\sin\theta_W\right].
\end{align}
The interaction with $Z$-boson and neutralino is somewhat
complicated. But based on the experience obtained above, some hints
could be obtained. From the expression of $y_{3/4}$, it can be
easily found that the $Z$-boson can be divided into two parts: one
is its transverse component proportion to
$(-N_{11}^*\sin\theta_W+N_{12}^*\cos\theta_W)$, and the other is the
longitude component proportional to
$(N_{13}^*\cos\beta-N_{14}^*\sin\beta)$. The interaction of
goldstini with the transverse part of the $Z$-boson is almost the
same as photon because they come from the same origin
$\frac{im_{a,i}}{\sqrt{2}F_i}\eta_i\sigma^{\mu\nu}\lambda^aF^a_{\mu\nu}$.
After multiplying a factor $\tilde{m}_1/m_{1}$ to
$N_{11}^*\cos\theta_W$ and a factor $\tilde{m}_2/m_2$ to
$N_{12}^*\sin\theta_W$ in Eqs.(\ref{ZG1},\ref{ZG2}), we can get the
coefficient of this transverse coupling, which is proportional
to $\tilde{m}_1$ or $\tilde{m}_2$.
For the longitude component of the
$Z$ boson, since it has some relations with the tilted mass
parameters of the Higgs, such as $\tilde{m}_{h_{u,d}}$ and
$\tilde{\mu}$, we can not get the same simple result. In
\cite{Argurio:2011gu} a free factor $K_{Z_L}$ is introduced to
connect the longitudinal interaction between $G$ and $G^\prime$.
Although we do not know the exact formula for $K_{Z_L}$, it must be
a function of $\tilde{m}_\phi/m_\phi$ from naive arguments and
dimensional analysis (here ${m_\phi}$ denotes the soft Higgs
parameters). The details of the coupling between $Z$-boson and
goldstini can be found in the appendix of \cite{Thaler:2011me}.

From the above analysis we see that the $G^\prime-\gamma-\chi$
and $G^\prime-Z-\chi$ (transverse $Z$) couplings are
both proportional to $\tilde{m}_a$ (tilted gaugino masses)
while the $G^\prime-Z-\chi$ (longitudinal $Z$)
and $G^\prime-h-\chi$ couplings are proportional to
$\tilde{m}_{\phi}$. So, if  $\tilde{m}_a$ are very small while
$\tilde{m}_\phi/m_{\phi}$ is much larger than one, we can get a new
scenario in which the Higgs decay $h\to \chi G'\to ZG'G'$ is more
sizable than
  $h\to \chi G'\to \gamma G'G'$.
Note that from the viewpoint of model-building, this scenario could be
easily realized in a two-sector messenger with $F_1\gg F_2$.
If the second sector preserves $R$-symmetry, it will give no
contribution to the gaugino masses.
Besides, in some concrete GMSB models, especially the
direct gauge mediation \cite{Komargodski:2009jf},
 the gaugino masses are usually suppressed by a factor $F_i^2/M_i^4$
due to the vacuum structure of superfields. So it is common to have
approximately vanishing $\tilde{m}_a$. Additionally, in the case of
stimulated SUSY breaking \cite{McCullough:2010wf}, it is quite
natural to split $F_1$ and $F_2$.

\section{Higgs decay to goldstini at the LHC}
From the analysis in the preceding section,
we see that in some GMSB models
the goldstini couplings  $G^\prime-h-\chi$  and  $G^\prime-Z-\chi$
can be sizable while the coupling $G^\prime-\gamma-\chi$ can be
much suppressed. This scenario can lead to the
Higgs decay $h\to \chi G'\to ZG'G'$ which will be studied in
this section.

\subsection{The rate of Higgs decay to goldstini}
We take an effective way to study the Higgs decay $h\to \chi G'\to ZG'G'$.
The effective Lagrangian is given by
\begin{align} \mathcal
{L}_{eff}=\frac{m^2}{F}[g_{h\chi}h\chi G^\prime + g_{\chi Z}
\bar{G}^\prime\bar{\sigma}^{\mu}Z_{\mu}\chi + h.c.] \label{eff}
\end{align}
Here a mass parameter $m$ is introduced to make the couplings
$g_{h\chi}$ and $g_{\chi Z}$ dimensionless and the Lagrangian takes
the similar form as in \cite{monophoton} which studied the signal of
mono-photon plus missing energy from the Higgs decay. Note that for
$m_{\chi}>m_h$, the lightest neutralino can decay to the Higgs boson
which may lead to many boosted Higgs bosons at the LHC
\cite{Howe:2012xe}.

\begin{figure}[htbp]
\includegraphics[width=3.2in,height=2.4in]{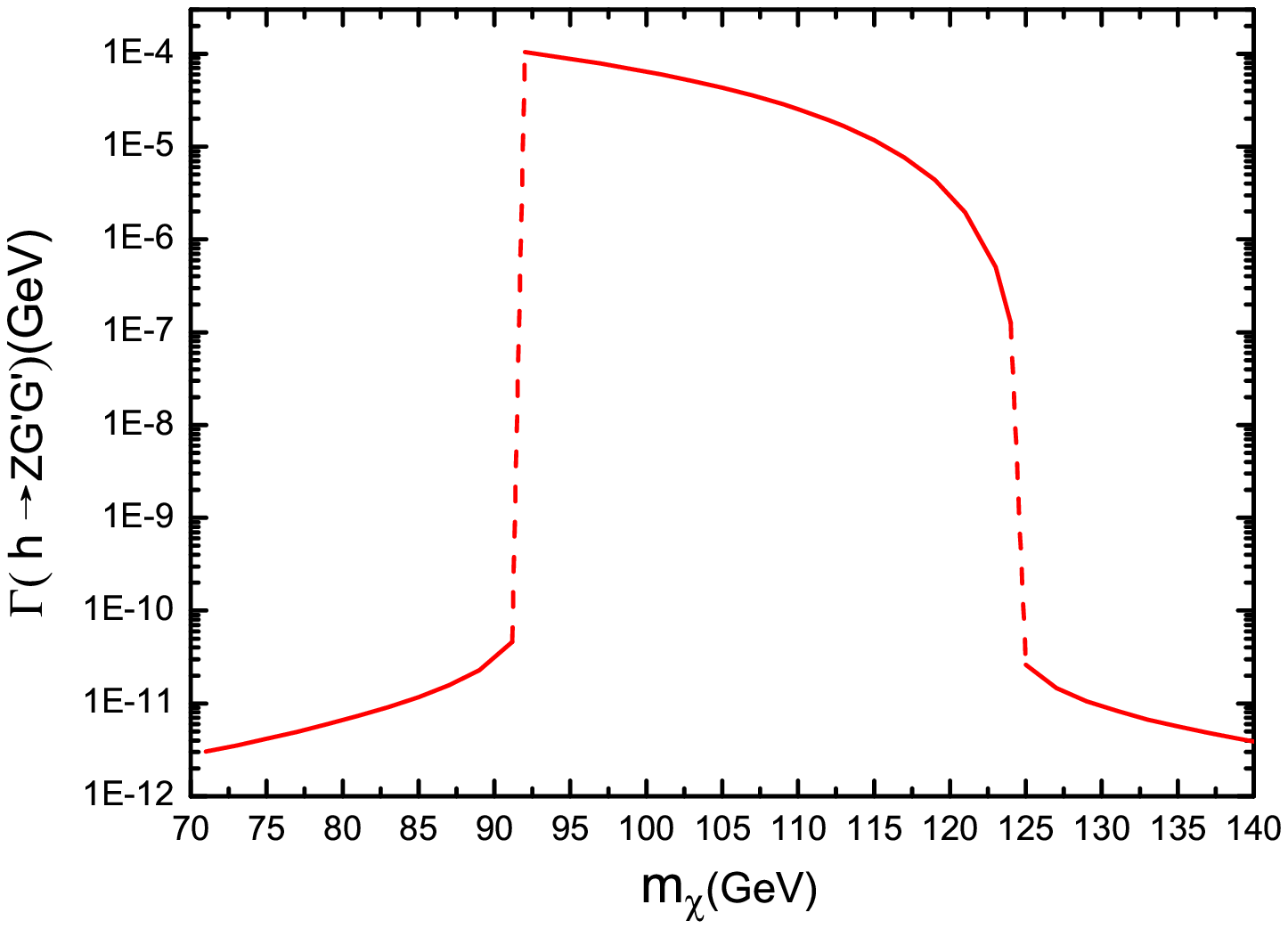}%
\hspace{-0.25in}%
\includegraphics[width=3.2in,height=2.4in]{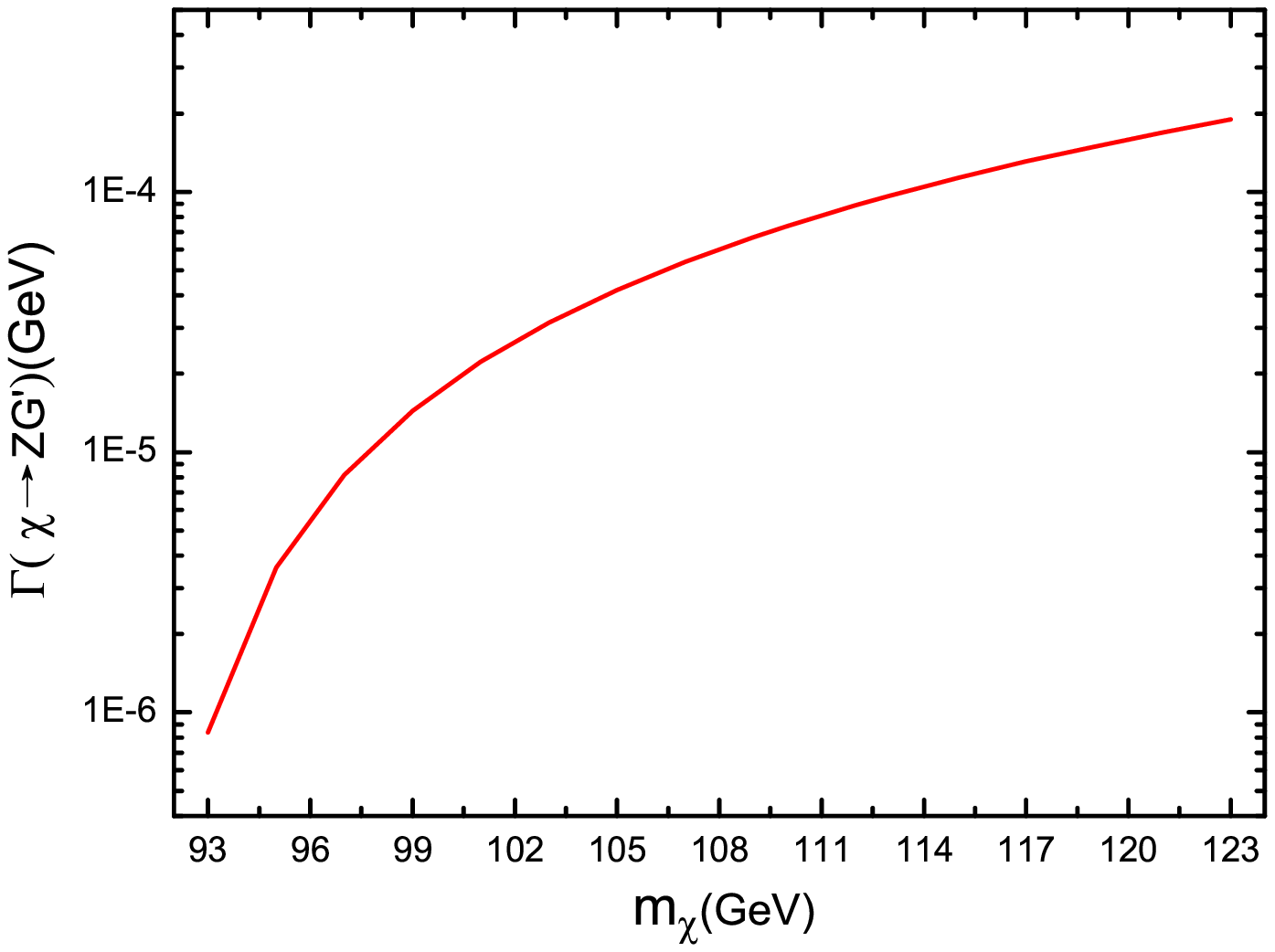}%
\hspace{0in}%
\vspace{-0.5cm} \caption{The partial widths of  $h\to ZG'G'$ and  $
\chi\to ZG'$ versus the neutralino mass. Here we choose $m =
0.1\sqrt{F}$ with $\sqrt{F} = 1.5$ TeV, and $g_{h\chi}=g_{\chi
Z}=1$.} \label{fig1}
\end{figure}

As shown in Fig.~\ref{fig1}, the decay branching ratio of $h\to ZG'G'$
is very small for an off-shell neutralino ($h\to \chi^* G'\to ZG'G'$)
but can be sizable for an on-shell neutralino ($h\to \chi G'$ followed by
$\chi \to ZG'$). The partial decay width of $h\to \chi G'$ is given by
\begin{eqnarray}
\Gamma(h \to \chi G')
&=&\frac{m_{h}}{8\pi}\frac{g_{h\chi}^{2}m^{4}}{F^{2}}\left(
1-\frac{m_{\chi}^{2}}{m_{h}^{2}}\right) ^{2}.
\end{eqnarray}

Note that there are already some constraints on the whole SUSY
breaking scale $\sqrt{F}$ through the interaction of gravitino
(usually photon is also involved). The current lower bound is around
300 GeV \cite{Acosta:2002eq,Achard:2003tx} and it is expected that
the LHC could push it to about 1.6 TeV \cite{Brignole:1998me}. The
most sensitive process to $\sqrt{F}$ is the decay channel
$h\rightarrow \chi G^\prime$ whose branching ratio is proportional
to $1/F$. In our calculation we fix $\sqrt{F}=1.5$ TeV and
$m=0.1\sqrt{F}$, and assume all the dimensionless couplings to be
unity. Note that here we only take a conservative value for the
parameter $m$. Since it is proportional to the tilted soft
mass, as discussed in the preceding section, it can be much larger.

About the decay of the neutralino $\chi$, it may have
some other modes, e.g., decay to SM particles if R-parity
is violated or decay to a light $U(1)_X$ gaugino \cite{Baryakhtar:2012rz}.
In our calculation we assume that $\chi$ only decays to $ZG^\prime$.
If there are other decay modes, the corresponding signal
should be multiplied by the branching ratio $Br(\chi\rightarrow ZG^\prime)$.
Additionally, the decay length of the neutralino with energy E is
approximately
$\Gamma^{-1} \sqrt{(E^2-m_\chi^2)/m_\chi^2}\sim
10^{-10}cm$, so the neutralino will decay inside the detector.

\subsection {Signal and background}
\begin{figure}[htb]
\epsfig{file=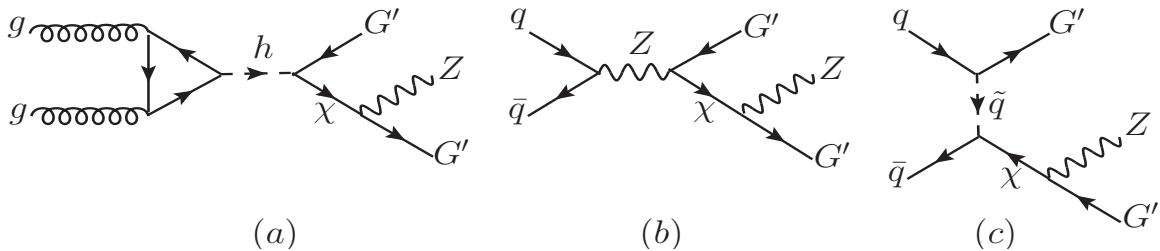,width=16cm,height=3.5cm} \vspace{0.25cm}
\caption{Feynman diagrams for the production of $ZG'G'$ at the LHC
induced by the goldstini couplings in Eq.~(\ref{eff}).}
\label{fig2}
\end{figure}

With the goldstini couplings in Eq.(\ref{eff}), the final state
$ZG'G'$ can come from three processes at the LHC, as shown in
Fig.~\ref{fig2}. Generally, the processes (a) and (b) will generate
$Z$-boson with low transverse momentum, while (c) with $t$-channel
squark exchange may generate a $Z$-boson with large transverse
momentum \cite{Brignole:1998me}. Additionally, heavier squarks which
are consistent with the result of LHC will suppress the $t$-channel
contributions, so we will not consider the process (c).

We require the $Z$-boson produced in association with goldstini to
decay to electrons or muons. Therefore, the signal is a pair of
electrons or muons and low missing transverse momentum, i.e., $pp
\to Z + {\large \not} \! E_{t} \to l^{+}l^{-} + {\large \not} \!
E_{t}$ ($l = e,\mu$). Some studies about such mono-Z process have
been performed in \cite{invisible
z'decay,mono-z-dm,Carpenter:2012rg}, where the mono-Z has a large
transverse momentum. In our signal, the $Z$-boson has small
transverse momentum because it comes from the decay of the
neutralino.

Note that the direct neutralino pair production $pp\to \chi\chi$
followed by the decay $\chi\to Z G^{\prime}$ with
one $Z$-boson decay to leptons and the other to neutrinos
can also give the final state  $l^{+}l^{-} + {\large \not} \! E_{t}$.
We checked that the rate of this process
is much smaller than the $s$-channel processes shown in Fig.2.
So we can neglect it safely.

For the $s$-channel process shown in Fig.2(a),
the Higgs boson can be produced on-shell and then decay to the
$Z$-boson plus missing energy. For such a production we can
get the high order corrections by multiplying a $k$-factor.
At the LHC, the leading order Higgs
boson production in the SM is from the gluon-gluon fusion. We
calculate the Higgs boson production cross section at $\sqrt{s}=14$
TeV with CTEQ6M parton distribution functions. We set the
renormalization scale $\mu_{R}$ and factorization scale $\mu_{F}$ at
the Higgs mass of 125 GeV. The other relevant parameters
are set as
\begin{eqnarray}
m_{t}=173.3 ~{\rm GeV},  ~~m_{b}=4.67 ~{\rm GeV},
~~m_{Z}=91.188 ~{\rm GeV},~~m_{W}=80.4 ~{\rm GeV}
\end{eqnarray}
We consider the higher order corrections and follow
\cite{monophoton} to get $\sigma_h^{NNLO}=49.99$~pb at the
next-to-next-to leading order with next-to-next-to leading logarithm
resummation. For the $Z$-boson $s$-channel process shown in
Fig.2(b), the neutralino can be produced on-shell and then decay to
$Z$-boson and $G^\prime$. So far no high order corrections have been
calculated for this process and we calculate this production at the
leading order.

The main SM backgrounds are
\begin{eqnarray}
 pp &\rightarrow& Z Z \rightarrow l^{+}l^{-}\nu\bar{\nu} , \\
 pp &\rightarrow& W^{+} W^{-} \rightarrow l^{+}\nu  l^{-}\bar{\nu} , \\
  pp &\rightarrow& Z j \rightarrow l^{+}l^{-} j ,\\
  pp &\rightarrow& t \bar{t} \rightarrow
 b\bar{b}l^{+}l^{-}\nu\bar{\nu} ,\\
 pp &\rightarrow& Z W^{\pm} \rightarrow l^{+}l^{-}l^{\pm}\nu .
\end{eqnarray}
Obviously, the first two processes are our irreducible backgrounds.
Because they proceed through the electroweak interaction, their
cross sections are expected to be relatively small at the LHC. The
third process can mimic our signal when the jet is missing detection,
which is detector-dependent. For the forth process to
mimic our signal, the two $b$-jets must be missing, which is less
likely. However, since the $t\bar{t}$ production rate is large
at the LHC, we cannot ignore it. For the last process, one
lepton must be missing. Since its
cross section is much smaller than the third process,
we neglect it in our calculation.

In our calculation we use MadGraph5 \cite{MG5} for both the signal
and backgrounds. For $Z$+jet process, we carry out parton shower
by Pythia \cite{Pythia} and match the matrix element with parton
shower in the $k_T$-jet MLM scheme \cite{MLM-scheme}. We perform
a fast detector simulation by using Delphes \cite{Delphes} and
use the anti-$k_t$ algorithm \cite{anti-kt} with the radius parameter
$\Delta R = 0.6$ to cluster jets.

\subsection {Numerical results}
Before giving the numerical result on the signal and backgrounds, we
summarize our selection criteria on the final state:
\begin{eqnarray}
&P_{T}^{l} > 20~{\rm GeV},~~{\large \not} \! E_{T} > 20~{\rm GeV},
~~|\eta_{l}| < 2.5,
~~\Delta R \equiv \sqrt{\Delta \eta^{2}+\Delta \phi^{2}} > 0.4,& \nonumber \\
&{\rm no~jet~with~} P_{T}^{j} > 20~{\rm GeV} ~{\rm and}~  |\eta_{j}| < 4.0,&
\label{criteria}
\end{eqnarray}
where the separation $\Delta R$ is for the leptons in the final
state with $\Delta \eta$ being the pseudo-rapidity difference and
$\Delta \phi$ being the azimuthal angle difference.
The jet veto and the lower cut on  the missing energy
can greatly suppress the $Z$+jet and $t\bar{t}$ backgrounds.
We further require the invariant mass of the dileptons to peak at
$Z$-boson mass and apply an upper cut of 30 GeV on the missing energy
(for the main signal process $pp \to h \to Z +
{\large \not} \! E_{t}$, the missing transverse momentum
is below about 30 GeV).

In Table I we present the number of events for the LHC
with $\sqrt{s} = 14$ TeV and 10
$fb^{-1}$ of integrated luminosity. In our calculation of the
signal, the mass parameters in Eq.~(\ref{eff}) are fixed as
$m_{\chi}=110$ GeV, $m_{G^{\prime}}=0$ (the goldstini is much
lighter than the Higgs, so we set it to zero for simplicity),
 $m=0.1\sqrt{F}$,
$\sqrt{F}=1.5$ TeV, while all the dimensionless couplings are fixed
to unity.
From this table we see that the signal is overwhelmed by
the backgrounds. As expected, the cut on the invariant dilepton mass
$|m_{ll}-m_{Z}|<5$ GeV can suppress the $W^{+}W^{-}$ and  $t\bar{t}$
 backgrounds efficiently.
The upper limit cut on the missing transverse momentum
can suppress the $ZZ$, $W^+W^-$ and  $t\bar{t}$ processes
significantly. But the $Z$+jet process cannot be suppressed by
this cut because here the missing jet has a low  transverse
momentum \cite{mono-z-dm}.

\begin{table}
\caption{The number of signal and background events
 for the LHC with $\sqrt{s} = 14$ TeV and 10 $fb^{-1}$
of integrated luminosity.} \vspace{0.2cm}
\begin{tabular}{||c||c|c|c|c|c||c|c|c||}
 \hline \hline
 \multirow{2}{*}{\minitab{c}{$\sqrt{s} = 14$ TeV \\($10fb^{-1}$)}}  &  \multicolumn{5}{c||}{Background (B)}  &  \multicolumn{3}{c||}{Signal (S)}  \\
\cline{2-9}
 &~~~$Z Z$~~~  &  ~~$W^{+} W^{-} $~~  &  ~~~$t\bar{t}$~~~  &  ~~~~$Z j$~~~~   &
 ~~~~$B_{tot}$~~~~  &  ~~$S_{h}$~~  &  ~~$S_{Z}$~~  &  ~~$S_{tot}$~~ \\
 \hline
 selection criteria &  964  &   12598  &  94  &  4812  &  18468  &  31  &  1.9  &  32.9 \\
 \hline
 $|m_{ll}-m_{Z}|<5$ GeV   &  834  &  900  &  35   &  3609  &  5378   &  25   &  1.7  &  26.7  \\
\hline
 ${\large \not} \! E_{T} < 30$ GeV  &  133  &  214  &  0   &  3609  &  3956   &  25   &  0.5  &  25.5  \\
 \hline
 \hline
\end{tabular}
\label{table 1}
\end{table}

Finally, we show in Table II the signal significance for different
integrated luminosity. Since the contribution from the $s$-channel
$Z$-boson process is much smaller than the Higgs process (as shown
in Table \ref{table 1}), in Table II we only consider the Higgs
process so that the result could be interpreted as the product of
the Higgs production rate, the Higgs decay branching ratio and the
neutralino decay branching ratio.

Note that in Table II we only show the statistical significance.
Since the ratio of signal to background
is quite small (around 0.6 percent), we need a quite high luminosity
to get a good statistic significance.
However, the detection of such a rare process also needs a good understanding
of the systematical uncertainty of the backgrounds, which is always a challenging
job for hadron colliders like the LHC. These backgrounds (especially the Z+jets)
are also the backgrounds for the Higgs signal $ZZ^*$ and so far at the LHC their
 systematical uncertainty is at a few percent level, as shown in Table 6 in
\cite{zz-atlas}.
In the future high luminosity 14 TeV LHC, the systematics of such backgrounds
need to be further improved for precision measurement of the Higgs property and
the probe of new physics.
Theoretically, we can have a larger ratio of signal to background
if we enlarge the effective mass parameter $m$ (in our above analysis we took a
conservative value of $0.1\sqrt{F}$ for illustration).

\begin{table}
\caption{Same as Table I, but showing the event number of the signal
$pp \to h \to \chi G' \to Z G' G' \to l^{+}l^{-} + {\large \not} \! E_{t}$
and its statistical significance for the LHC with $\sqrt{S} = 14$ TeV and different
luminosity. } \vspace{0.2cm}
\begin{tabular}{||c|c|c|c|c|c|c||}
 \hline \hline
 ~~~$\sqrt{S} = 14$ TeV~~~  &  ~100$fb^{-1}$~  &  ~500$fb^{-1}$~  &  ~800 $fb^{-1}$~  &  1000 $fb^{-1}$  &  2000 $fb^{-1}$  &  3000 $fb^{-1}$  \\
 \hline
 $S_{[selection~criteria]}$ &  310  &  1550  &  2480   &  3100  &   6200   &  9300  \\
 \hline
 $S_{[passing~all~cut]}$  &  250  &  1250  &  2000  &  2500  &  5000   &  7500  \\
\hline
 $S$/$\sqrt{ S+B}$  &  1.3  &  2.8  &  3.5  &  4.0  &  5.6  &  6.8  \\
 \hline
 \hline
\end{tabular}
\label{table 2}
\end{table}

\section{Conclusion}
Compared with the gravitino, the goldstini can couple more strongly
to the visible fields and thus induce some interesting
phenomenology. In this note we considered the effects of the
goldstini on the Higgs phenomenology. We found that in some GMSB
models the goldstini has approximately vanishing interaction with
photon and the lightest neutralino, but the corresponding coupling
with the $Z$-boson is sizable. This could induce the mono-$Z$ decay
of the Higgs boson ($h\to G^\prime+\chi \to Z+2G^\prime$) which
signals $Z$-boson and ${\large \not} \! E_{t}$ at the LHC. Then in an
effective model with conservative fixed parameters we studied the
observability of this decay at the LHC. From Monte Carlo simulation
of the signal and backgrounds, we found that it is not accessible at
the finished 8 TeV run (25 fb$^{-1}$) or 14 TeV run with 100
fb$^{-1}$, but might be observed at the high luminosity LHC
(14 TeV, 1000-3000 fb$^{-1}$) if the systematics of the backgrounds can
be well understood. Although at the LHC it is so
challenging to detect this exotic decay channel of the Higgs boson,
it is worth hunting because such a scenario may naturally exist in
GMSB with multi SUSY breaking sectors.

\section*{Acknowledgments}
We would like to thank Olivier Mattelaer for his great help on
MadGraph and Lei Wu for useful discussions. Lin Wang acknowledges
Prof.~Johann H. K\"{u}hn and Prof.~Matthias Steinhauser for their
warm hospitality. This work is supported by DFG through SFB/TR 9
``Computational Particle Physics'' and by the National Natural
Science Foundation of China under grant Nos. 11275245, 10821504 and
11135003.

\end{document}